\documentclass[12pt,letterpaper]{article}
\usepackage{amsmath,amssymb,pgf,pgfarrows,pgfnodes,float,appendix, hyperref}
\usepackage{graphicx}
\usepackage{subfigure}
\usepackage[margin=0.9in]{geometry}

\newcommand{\be}{\begin{equation}}
\newcommand{\ee}{\end{equation}}
\newcommand{\bea}{\begin{eqnarray}}
\newcommand{\eea}{\end{eqnarray}}

\title{{\rm\footnotesize \qquad \qquad \qquad \qquad \qquad \ \qquad \qquad \qquad \ \ \ \ \ \                 RUNHETC-2023-35}\vskip.5in   Old Ideas For New Physicists: II}
\author{Tom Banks\\
Department of Physics and NHETC\\
Rutgers University, Piscataway, NJ 08854\\
E-mail: \href{mailto:tibanks@ucsc.edu}{tibanks@ucsc.edu}
\\
\\
\\
\\
}
\date{}
\begin{document}
\maketitle

\begin{abstract}  We review ideas relating the breaking of Supersymmetry in de Sitter (dS) space-time to amplitudes in which gravitinos are reflected from the dS horizon.  A self consistent evaluation of such amplitudes leads to the estimate $m_{3/2} = K \Lambda^{1/4}$, with a constant $K$ that is not easily calculable.  LHC bounds on superpartner masses, plus the structure of low energy effective field theory then imply that there must be new superfields with standard model quantum numbers in the region below $100$ TeV.  Only one class of models, the Pyramid Schemes, has been found, which is compatible with one loop coupling unification.  They have a new strongly coupled gauge sector with a confinement scale in the multi TeV region. We outline the general predictions of these models, which include Dirac gauginos and a novel solution of the strong CP problem.  The models all appear to have a "little hierarchy problem": the expectation value of the standard model Higgs field, and its mass, must be a factor of $10$ or more smaller than the masses of the second Higgs doublet or standard model superpartners.  Since all of these models contain uncertainties from the strong coupling physics, as well as the physics of the dS horizon, it is difficult to determine how serious this fine tuning problem is.   \end{abstract}
\maketitle

\section{Introduction}

All known consistent perturbative string compactifications to asymptotically flat space have exact Supersymmetry (SUSY).  With one class of exotic exceptions\cite{evaetal}, all well understood families of $AdS/CFT$ models with a zero c.c. limit, are exactly SUSic.  The real world, on the other hand, is not SUSic and appears to have a very small positive cosmological constant (c.c.).   Neither perturbative string theory nor AdS/CFT applies to it directly.  In this note I will review some old ideas about the relation between the breaking of SUSY in the real world and the value of the c.c..  In a nutshell, the claim is that SUSic solutions of SUGRA generally have negative c.c. and zero c.c. models are characterized by the existence of a discrete chiral R symmetry, which sets the gravitino mass to zero.  Note that in SUSic models in AdS space, the gravitino is in a superconformal multiplet with the graviton and does not have zero mass in effective field theory.  The breaking of SUSY is intimately connected with the breaking of this chiral R symmetry by reflection of gravitinos from the cosmological horizon.

In\cite{oldideas1}\cite{hilbertbundles} we explored the argument that de Sitter (dS) space was a system with a finite dimensional Hilbert space, whose holographic degrees of freedom were encoded in copies of the cutoff spinor bundle on the finite area holoscreen of the cosmological horizon.  Chiral symmetries are always broken by the boundary conditions on space-times with boundaries.  If we view an asymptotically flat SUSic model as the limit of a model with positive c.c. then the breaking of R symmetry should be associated with interactions in which the gravitino, or other R charged particle, is reflected from the horizon.   Any such amplitude will have the form
\begin{equation} e^{- 2 m_{3/2} R} {\rm Tr} \rho_h V V^{\dagger} , \end{equation} where $\rho_h$ is the horizon density matrix and $V$ the operator that causes the horizon to absorb one gravitino.  We've taken into account the exponential falloff of two massive propagators in dS space, but not their power law dependence.  A gravitino of mass $m_{3/2}$ can remain within a Planck length of the horizon for a proper time of order $m_{3/2}^{-1}$.  During that time, if it performs a random walk over the holographic screen, it can cover an area of order $C (m_{3/2} M_P)^{-1}$, if the length of a step is $L_P$.  Using the area law for entropy, the number of intermediate states that can contribute to the operator product $VV^{\dagger}$ is $e^{C  (m_{3/2} / M_P)^{-1}}$.   This means that the value of the gravitino mass is of order
\begin{equation} m_{3/2} \sim e^{- 2 m_{3/2} R} e^{C  (m_{3/2}/M_P)^{-1}} . \end{equation}
For consistency, $m_{3/2}$ has to go to zero as $R$ goes to infinity.  However, if it goes to zero too rapidly, the second factor on the right hand side of this estimate blows up.  If it goes to zero so slowly that the first factor dominates, then the estimate predicts that it vanishes exponentially, which is again a contradiction.  The only consistent behavior is for the two exponentials to cancel, allowing corrections to the estimate to control the falloff.  This implies
\begin{equation} 2 m_{3/2} R = C  (m_{3/2}/M_P)^{- 1} . \end{equation}  Thus
\begin{equation} (m_{3/2}/M_P) =  B (RM_P)^{- \frac{1}{2}} , \end{equation} where \begin{equation} B = (\frac{C}{2})^{\frac{1}{2}} . \end{equation} 

As we'll note below, standard supergravity formulae then imply a SUSY breaking scale below $100$ TeV.  Thus we expect a renormalizable effective field theory at the multiple TeV scale to describe SUSY breaking.  It is impossible to construct such a model, which is consistent with experiment, using just standard model singlet fields and the SUSic standard model.  Thus the combination of the idea that SUSY breaking originates from gravitino reflection off the cosmological horizon, with effective field theory, requires us to add a new sector to effective field theory at energies close to the TeV scale.   The new sector must contain fields charged under the standard model gauge group, and making it compatible with gauge coupling unification puts major restrictions on what it can be.  So far only one class of models has been found that satisfies all the constraints.  It is called the Pyramid Scheme, because of the shape of its quiver diagram.  

Our novel perspective on SUSY breaking also turns out to have interesting implications for the strong CP problem.   It turns out that in the limit in which the c.c. vanishes, the model has an exact symmetry, which guarantees that the only CP violating angle in the low energy effective theory is the CKM angle.  This symmetry is broken by the same gravitino diagrams that generate the gravitino mass, but one can give plausible arguments that the CP symmetry breaking terms are too small to generate a measureable strong CP angle.  

\section{A New Strongly Interacting Sector Close to the TeV Scale}

Let us begin by recalling the estimate for the scale of Cosmological SUSY breaking.  In the limit of vanishing c.c. the gravitino mass is protected by a discrete $Z_N$ R symmetry. The gravitino mass is generated by diagrams in which the gravitino reflects off the horizon.  These have the form
\begin{equation} e^{- 2 m_{3/2} R} \sum \langle s | T | n \rangle\langle n | T^{\dagger} | s' \rangle . \end{equation}  Here we've assumed that the horizon is initially in a pure state .  In a Feynman diagram calculation the gravitino line strikes the horizon at some point and then performs a random walk over the horizon.  Since, in a self consistent calculation, it is massive, the proper time of that walk is of order $m_{3/2}^{-1}$ and so it covers an area of order $m_{3/2}^{-1} L_P$ .  Here we've assumed that the length of a step in the walk is the Planck length.  The Gibbons-Hawking area law tells us that the number of states contributing positively in the sum is of order $e^{C (L_P m_{3/2})^{-1}}$.   If we now imagine that $m_{3/2}$ vanishes like some power of $R$ as $R/L_P \rightarrow \infty$ then the only consistent scaling
is
\begin{equation} m_{3/2}^2 = (C/2) (RL_P)^{-1} = (C/2) 10^{-23} ({\rm GeV})^2 = [\sqrt{C/2} 10^{-2.5} {\rm eV}]^2 .\end{equation}

The standard supergravity (SUGRA) formula for the gravitino mass is
\begin{equation} m_{3/2} (1 {\rm eV})^{-1} = \sqrt{F} (65 {\rm TeV})^{-1} . \end{equation}
So our formula predicts 
\begin{equation} \sqrt{F} = .65 {\rm TeV} \sqrt{\frac{C}{20}} . \end{equation}  So one needs a fairly large value of $C$ to make this idea compatible with experiment.  Our understanding of the physics of the dS horizon, and of the way in which effective field theory emerges as an approximation to the correct theory of quantum gravity in dS space, is so primitive that one can easily imagine that factors of $100$ could be missing from our estimates.  In addition we know that there is at least one other large energy scale, the scale of coupling unification, evident in known physics.  Neutrino masses, and many of the models invoked to explain the hierarchies of quark and lepton Yukawa couplings, suggest that the region $10^{14} - 10^{16}$ GeV may have quite a bit of structure in it.  All of this could modify our crude estimate of the gravitino mass.

If the SUSY breaking scale is this low, and there are no fields transforming under the standard model gauge group in effective field theory below 100 TeV, besides those of the minimal  Supersymmetric Standard Model (MSSM), then it is impossible to get a superpartner spectrum consistent with experiment.  The only renormalizable coupling of singlets to the MSSM is some linear combination of singlet fields coupling to $H_uH_d$  and this can generate squark and gluino masses only through loops, which will be much too small.  

Once we add new low energy fields with standard model couplings, and a new strong interaction scale to justify their existence, we have to worry about coupling unification.  So far only two general classes of models which seemed to be consistent with coupling unification and the idea of low scale SUSY breaking have been found: the Pentagon models\cite{pentagon} and the Pyramid schemes\cite{pyramid}.  Detailed analysis\cite{jeff} shows that the Pentagon models don't work, while the Pyramid schemes may be viable.

\begin{figure}
\begin{center}
\includegraphics[scale = 0.4]{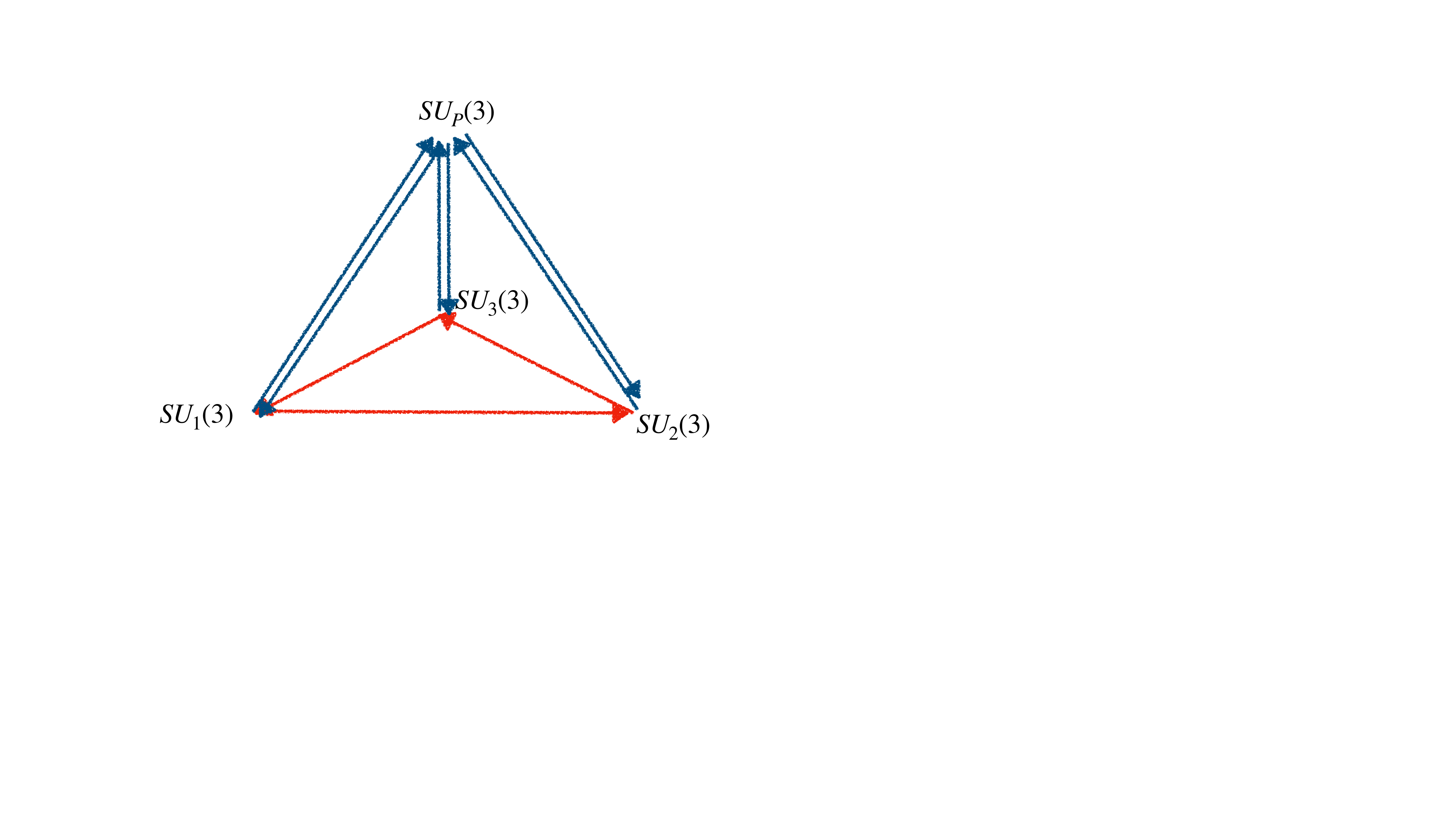}
\caption{The quiver diagram of the Pyramid Schemes. Trianon fields connect the apex of the Pyramid to the Trinification model of Glashow and Georgi at the Base.  Different versions of the Scheme have different Singlet couplings to this quiver.}
\label{fig:quiverofpyramid} 
\end{center}
\end{figure}

The quiver diagram of the Pyramid Schemes \ref{fig:quiverofpyramid} is a three sided pyramid whose base is the trinification model of Georgi and Glashow\cite{trinification}.  

This unified model has gauge group $SU(3)_3 \times SU(3)_2 \times SU(3)_1 \ltimes Z_3$, where the $Z_3$ cyclicly permutes the three $SU(3)$ groups.  Each field connecting the apex of the pyramid to the base is also a fundamental + anti fundamental of the group at the apex of the pyramid, which is also $SU(3)$.    We call it $SU_P (3)$.

Color is the $SU_3 (3)$ group of the base of the pyramid, and the $SU(2)$ of the electroweak interactions is a Cartesian $SU(2)$ in $SU_2 (3)$.  Weak hypercharge is a linear combination of the hypercharge generators of $SU_2 (3)$ and $SU_1 (3)$.  The 15 Weyl fermions of the standard model are part of the $[27]$ dimensional representation of the base of the pyramid and singlets under the apex.  The other fields in the $[27]$ have masses of order $10^{16}$ GeV and their absence at low energy gives rise to the running of the couplings.  Some of these fields might actually be the Higgs fields of the MSSM, but in the effective Lagrangian below the unification scale, to one loop order, we do not have to commit ourselves to such an identification.  

The fields connecting the apex and the base are called {\it trianons} and form complete multiplets of the trinification group at the base.  As long as they are degenerate, they do not effect the one loop running of standard model couplings. 
The apex $SU_P(3)$ thus has $9$ flavors of  $[3] + [\bar{3}]$ chiral superfields.  We can organize these fields into six $3 \times 3 $ matrices $T_i$ and $\bar{T_i}$ where one index is the apex gauge quantum number and the other labels the gauge quantum number under the $i$-th $SU(3)$ of the base of the pyramid.  The unique gauge invariant superpotential of dimension $\leq 4$ made from these fields is 
\begin{equation} \sum_i [u_i {\rm det}(T_i) + \bar{u}_i {\rm det} (\bar{T}_i) ]  . \end{equation}

In addition to the supersymmetric standard model, we will introduce $P$ singlet fields $S_p$ whose role is to convey SUSY breaking to the rest of the system.  In the limit of vanishing c.c., the system has a discrete symmetry group, one factor of which acts on the supercharges and guarantees that the c.c. is zero, a fact which is otherwise unnatural in SUGRA even when SUSY is preserved.  We have a superpotential

\begin{equation} W_{\Lambda = 0} = \sum_{pj} y_{pj} S_p {\rm tr} T_j \bar{T}_j + \sum_i [u_i {\rm det}(T_i) + \bar{u}_i {\rm det} (\bar{T}_i) + \beta_p S_p H_u H_d] + Y^u_{IJ} H_u \bar{U}_I Q_J + Y^d_{IJ} H_d \bar{D}_I Q_J + Y^l_I \bar{E}_I H_d L_J  . \end{equation} The discrete symmetry group is chosen so that no other terms of dimension $4$ appear in the superpotential and that the only terms of dimension $5$ that appear in the effective action, and which violate baryon or lepton number, are those that give rise to neutrino masses
\begin{equation} W_{\nu} = \frac{G_{IJ}}{M} (L _I H_u)(L _J H_u) . \end{equation}
All higher dimension operators are presumed to be scaled by the unification scale $10^{16}$ GeV.  $M$ should be slightly smaller and the explanation of that is left for unification scale physics, as is the structure of the Yukawa couplings.  

In\cite{kathrein} it was shown that this can all be accomplished by a $Z_8$ R symmetry with R charge assignments
\begin{equation} S_i = T_i = \bar{T_i} = 6, \ \ \ \ Q_I = 5, \ \ \ \ L_I = 1, \ \ \ \ H_d = 0, \end{equation} along with
\begin{equation} H_u = - 4, \ \ \ \ \bar{U}_I = 1, \ \ \ \ \bar{D}_I = - 3, \ \ \ \ \bar{E}_I = 1 . \end{equation}
In addition, we must invoke an ordinary discrete symmetry to forbid cubic terms in the singlet fields.

\subsection{Breaking the Discrete Symmetries}

We now want to add explicit breaking of the discrete symmetries to mimic the effects of the dS horizon.  These should enforce the relation
\begin{equation} m_{3/2} = K \Lambda^{1/4} , \end{equation} which we derived by considering self consistent gravitino mass generation via reflection off the horizon.   There is no reason for them to obey the genericity constraints of traditional low energy effective field theory.    We choose
\begin{equation} \delta W = W_0 + \sum [m_i {\rm tr}\ (T_i \bar{T}_i) + \mu_p^2 S_p]  + \mu H_u H_d . \end{equation}  If we take $\mu_p^2 \sim \Lambda^{1/4} M_P $, $m_i \sim \Lambda^{1/8} M_P^{1/2}$,  and $W_0 \sim \Lambda^{1/4} M_P^2$ , then we can satisfy the relationship between the gravitino mass and the c.c. .  Note that since these terms are non-generic, we avoid the theorem of Nelson and Seiberg\cite{ns} about the necessity of R symmetry for spontaneous SUSY breaking.  The $F$ terms of the $S_p$ fields are non-vanishing at the potential minimum where the VEVs of all fields vanish.  Although this is not the global minimum of the potential, the fact that SUSY is restored in the limit that $\Lambda/m_P^4 \rightarrow 0$ implies that the Coleman De Lucia tunneling probability to a negative energy Big Crunch state will vanish exponentially with the dS entropy.  This is consistent\cite{tbjftunnel}\cite{abj} with an interpretation of this transition as a temporary sojourn into a low entropy state, as suggested by the Covariant Entropy Principle from a comparison of the maximal causal diamonds in the two space-times.

In treating $\delta W$ as part of an effective Lagrangian below the unification scale we must be aware of its origin in terms of diagrams with gravitinos propagating out to the horizon.  The local operators appearing in it are therefore "soft" and UV corrections to them much smaller than one would estimate by power counting.  This is why we don't have to worry about "large UV renormalizations of the c.c.".  The phrase in scare quotes makes no sense in terms of the underlying hypothesis that the c.c. is determined by the CEP, and it is avoided in EFT calculations by remembering the softness of $\delta W$.  This is somewhat analogous to what happens in technicolor models, where a mass term for a pseudo-goldstone boson is protected from large renormalization because the boson is composite.  

Another possible feature of $\delta W$ that is traceable to its unusual origin is a novel solution to the strong CP problem.  For $\Lambda = 0$ the Pyramid schemes incorporate the Peccei-Quinn solution to the strong CP problem.  We have written the most general Lagrangian, up to terms of dimension $4$ and $B$ conserving terms up to dimension $5$, consistent with a set of discrete symmetries.   It is easy to verify that all CP violating phases besides the QED theta angle and the CKM phase can be rotated away.   Let us assume that those phases are generated by physics somewhat below the unification scale.  An example might be the scale of $10^{14.5}$ GeV that appears to be associated with the dimension $5$ operator that gives rise to neutrino masses.  The term $\delta W$ is generated by interactions in which a gravitino is absorbed and re-emitted by the degrees of freedom on the cosmological horizon.  The reflected gravitino trajectory is highly accelerated and so experiences an Unruh temperature.  If that temperature is higher than the energy scale associated with CP violation, then the CP phases will average out and the parameters in $\delta W$ will be real.  The $\Lambda = 0$ Lagrangian has a Peccei-Quinn axion associated with an accidental $U(1)$ symmetry that is broken by $\delta W$, but if the Unruh temperature is higher than the CP violation scale, then $\delta W$ gives this axion a large mass without introducing new CP violating phases into low energy physics.  There is no "axion quality problem" in this variation of the PQ mechanism, because terms at the unifications scale, which might break the PQ symmetry are much smaller than those coming from $\delta W$.

\subsection{Approach to Phenomenology and Cosmology}

The original analysis of these models\cite{tbjf}\cite{kathrein} was done at a time before the Higgs boson was discovered, and during which a variety of intriguing indications for dark matter signals had been announced.   Most of those have gone away.  In addition, Fischler and the author have discovered a holographic model of inflation and dark matter\cite{holocosmrev} in which the dark matter particles are primordial black holes (PBHs) which carry the lowest value of a discrete $Z_N$ gauge symmetry charge.  They have mass of order $M_P = 10^{19}$ GeV.  This removes the necessity for the contortions of\cite{kathrein} which were necessary in order to make perturbative gauge coupling running compatible with the existence of a particle dark matter candidate.  The models of\cite{holocosmrev} give a plausible account of all features of the cosmology of the very early universe, and depend on the nature of low energy particle physics only through the number of species with masses below about $10^8$ GeV, and their spectrum.   Our analysis of the Pyramid schemes will concentrate only on particle physics.  The paper in earlier literature which we follow most closely is\cite{tbdiracgluino}.
We differ from that analysis mostly by insisting that all three pyrmabaryon number violating Yukawa couplings appear in the Lagrangian for $\Lambda = 0$.  This removes the bounds on the confinement scale $\Lambda_3$ from\cite{kathrein} as we recall below.  

We begin with a summary. $SU(3)$ SUSY gauge theory with $9$ flavors and the indicated Yukawa couplings (we ignore the couplings to the singlets for the moment), has an attractive line of fixed points at
\begin{equation} g_P^2 = \frac{3}{4} u^2 , \end{equation} where all three Yukawas are equal and $g_P$ is the gauge coupling.  This is a one loop result.   We assume that just below the unification scale the couplings are in the domain of attraction of this fixed line and flow to it at a point where the couplings are fairly strong.  {\bf It's not clear if we need to stay within the domain of validity of the one loop calculation}.  We do so only in order to have control of the calculation.

Since the trianons are in a full multiplet of the trinification group, the one loop standard model coupling ratios flow in a way that is unchanged, if we assume that the values just below the unification scale are perturbative, but the QCD coupling increases more slowly in the IR, because we have, in effect, 3 more quark flavors.  Thus, in order to match low energy physics we must start from a larger gauge coupling at the unification scale.  It is still consistent to assume that the one loop approximation is adequate. 
This pattern of RG flows continues until we hit the trianon mass thresholds.  The trianon masses all come from interactions with the horizon and we assume they're roughly equal.

\section{RG Flow and The Effective Theory Below $\Lambda_3$}

We begin with some unification scale values of all the gauge couplings and the Yukawa couplings in $W_{\Lambda = 0}$.  These are assumed to be such that one loop perturbation theory is a good approximation to RG running.  For the standard model couplings this means that the ratios run as they do in the MSSM, while the individual couplings are less asymptotically free because of the presence of the trianons.  $g_P$ and the Yukawa couplings of the trianons run rapidly to some point on the fixed line.

At some point we reach the scale of the trianon masses $m_{i}$ and $g_P$ begins to run towards strong coupling.  $SU_P (3)$ confines at a scale $\Lambda_3 $ below $m_{i}$.  
Generically, the point on the fixed line to which $g_P$ has run between the unification scale and the much lower scale $m_i$ is not weakly coupled.  Therefore, $\Lambda_3 \sim m_i$.   $SU_P (3)$ SUSY gauge theory with $9$ massive trianons may or may not have a meta-stable Poincare invariant state, which preserves the standard model gauge group and breaks SUSY, but we can certainly tune the couplings of the singlet fields $S_i$ to guarantee the existence of such a state, and then tune $W_0$ so that this state has a small positive c.c. in SUGRA, coinciding with the observed value.

It should be emphasized again that these tunings have a different character in models motivated by cosmological SUSY breaking than they would in conventional low energy effective field theory.  Conventionally, we are interested in the lowest minimum of the potential energy density.  In CSB we are modelling the properties of localized excitations of a dS model of quantum gravity.  The $\Lambda = 0$ limit of that model has excitations whose low energy scattering matrix (for small enough numbers of particles) can be computed in some particular $N = 1$ SUGRA model with a discrete R symmetry, which guarantees that the c.c. is zero in the low energy action.  It might be that the low energy action has a moduli space of SUSic, R symmetric solutions, but only one of them has anything to do with the $\Lambda \neq 0$ model from which we began.  In that model, interactions with the horizon generate a non-generic superpotential on the moduli space, including a constant $W_0$, which is tuned to make the c.c. small and positive {\it at one particular point in moduli space}.  This is the point which is supposed to correspond to the model of the real world.   Now consider the effective potential on moduli space and suppose it has a minimum lower than the one chosen to model the real world.   Then Coleman De Luccia instantons tunneling from the real world minimum to the "true minimum" are {\it above the Great Divide}\cite{abj}\cite{tbjftunnel}.  They correspond to temporary sojourns of the typical state of the finite entropy dS vacuum state into a highly improbable low entropy state.  The return probability is governed by the principle of detailed balance and is close to one.  Thus, in treating CSB with the tools of low energy field theory, we {\it must} choose minima of the effective potential whose symmetry properties correspond to those we observe, rather than search for a model where those properties emerge at the global minimum of the potential. 

For similar reasons we do not require the R breaking terms in the superpotential to be generic, as an effective field theorist would.  They are there to "retrofit" the EFT to an underlying model of quantum gravity whose structure is quite different from QFT.  For recent progress in understanding the relation, see\cite{hilbertbundles}.  We're still far from the point where we can compute the terms in the EFT from an underlying gravitational model.  We've learned from perturbative string theory that we can trust EFT symmetry arguments when $\Lambda = 0$, but we also know that those arguments are wildly misleading at the non-perturbative level.  Almost all of the perturbative Calabi-Yau compactifications of {\it e.g.} heterotic string theory with $N = 1$ SUSY, which are protected to all orders in string perturbation theory by non-renormalization theorems based on holomorphy and axion shift symmetry, do not really exist as non-perturbative models of quantum gravity.  Our philosophy in this paper will be to use conventional symmetry and genericity arguments for the $\Lambda = 0$ terms in the EFT, but to choose the terms that arise from "interactions with the horizon" to fit phenomenological considerations, without paying attention to EFT naturalness.  It's obvious that one must do this for the term $W_0$, which fixes the c.c..  In the introduction we've shown how the model of the origin of these terms from gravitino exchange might solve the strong CP problem.

\subsection{SUSY and Electroweak Symmetry Breaking}

Just above the scale $\Lambda_3$, our model is weakly coupled. If we assume that the squarks and sleptons don't get tachyonic mass terms via their perturbative couplings to the $SU_P (3)$ gauge theory, then the low energy effective potential is
\begin{equation} V = g_1^2 D_1^2 + g_2^2 D_2^2 + (|\beta_p S_p + \mu|^2) (|H_u|^2 + |H_d|^2) 
+ K^{p\bar{q}}(S/\Lambda_3 ,\bar{S}/\Lambda_3) (\mu_p^2 + \beta_p H_u H_d + \gamma_{pr} S_r)(\mu_p^{* 2} + \beta_p^* H_u^* H_d^* + \gamma_{pr}^* S_r^*) . \end{equation}   Recall that $\beta_p$ comes from the limiting model with $\Lambda = 0$ , while $\mu_p^2 $ and $\gamma_{pr}$ come from interactions with the horizon and vanish as $\Lambda/M_P^4 \rightarrow 0$.  The first two terms in the potential are the electroweak D terms.  They favor $\tan \beta = 1$.  Below $\Lambda_3$ we cannot compute the Kahler metric of the singlet fields, even if the couplings $\alpha_{pi}$ are small (as they should be if the model is to remain perturbative up to the unification scale), since it depends on correlation functions in the strongly coupled $SU_P (3)$ gauge theory.  However it is clear that, if we choose $\gamma_{pq}$ to have two zero eigenvalues, SUSY is broken. This implies that the number of singlet fields $P \geq 3$.

Before turning to a detailed discussion of constraints on the spectrum, let us first outline the amount of variation that we can expect in the various parameters, which would be determined by a more microscopic theory. The Kahler metric, perturbatively in the parameters $y_{pi}$ is determined by
\begin{equation}\sum_i y_{pi}y_{qi}^* \langle M_i M_i^* \rangle ,\end{equation} where $M_i$ is the singlet meson field constructed from the $i$-th trianon.  The parameters $y_{pi}$ come from some consistent model of quantum gravity in asymptotically flat space with $N = 1$ SUSY.  Perturbative string theory gives us continuous families of candidates, and F theory and M theory provide others.  Some of those have discrete R symmetries that guarantee that they will survive non-perturbative corrections\footnote{Most don't, despite being consistent to all orders in string perturbation theory because of axion shift symmetries.}.  There is no known catalog of these and there might be an infinite family of such consistent models.   One would then have to sift through those to find models with the right low energy gauge group structure to give a Pyramid scheme broken to the standard model times $SU_P (3)$, and determine how many low energy singlets they contained and the values of the couplings $y_{pi}$ and $\beta_p$.  While current string theory technology is not yet capable of performing this task, it is at least the type of task that traditional perturbative string theorists worked on for decades.  

The determination of the parameters $\mu_p^2 $ and $\gamma_{pq}$ from a more fundamental theory is more mysterious, and involves the poorly understood theory of de Sitter space.  According to the proposal of\cite{hilbertbundles}\cite{tbwfdS} this is a finite quantum system, which converges to the "string theory" S matrix in the limit of vanishing c.c..  The general mathematical theory of limits suggests that there will be many such systems, so that even given a choice of "target" SUSY S matrix, the values of the SUSY violating parameters have some allowed range and we are far from understanding what it is.   

All of the natural mass scales in our model, come from the multi TeV scale determined by CSB.  Even the confinement scale $\Lambda_3$ is close to this scale because we've assumed that the fixed line value to which the $SU_P (3)$ coupling is driven is strong, so that confinement sets in at the trianon mass scale.  Nonetheless, we are going to assume that the Higgs VEVs are somewhat smaller, so that $|H_u|^2 + |H_d|^2$ takes on the value $(250 {\rm GeV})^2$ in the vacuum.   This is the form of the "little hierarchy problem" in our model.  We will comment below on the amount of fine tuning this requires.
In the approximation that the Higgs VEVs are zero, the singlet field VEVS are determined by minimizing the effective potential 
\begin{equation} V_0 = K^{pq} (S,S*) (\mu_p^2 + \delta_p^3 \gamma S_3)(\mu_q^{*\ 2} + \delta_p^3 \gamma^* S^{*\ 3})  .   \end{equation}   SUSY is broken, the positive vacuum energy is canceled by the constant $W_0$ in the superpotential to give the observed valued of the c.c., and the singlet fields all have multi TeV scale VEVs.   The dimensionless Kahler metric has some fixed numerical value at the vacuum values of the scalars.   

Next we examine the quadratic terms in the Higgs fields, expanded around these singlet VEVS, $S_P^0$.   These have the form
\begin{equation} {\cal M}^2 H_u H_d + h.c.  + |\beta_p S_p^0 + \mu|^2 (|H_u|^2 + |H_d|^2),\end{equation} where $  {\cal M} $ is of order the multi-TeV scale .
The dynamics of the $SU_P (3)$ gauge theory preserves CP and we've argued that the horizon physics that determines the dimensionful parameters in $\delta W$ is also CP conserving, so ${\cal M}^2$ should be real.  It is easy to see that the potential is minimized when the phase of $H_u H_d$ is zero and the VEVs are oriented so the the $U(1)$ of electromagnetism is unbroken.   We are left with quadratic terms
\begin{equation} {\cal M}^2 H_u H_d + |\tilde{\mu}|^2 (H_u^2 + H_d^2) , \end{equation}  where the values of the field are real.   The first term is an anti-symmetric matrix which has equal positive and negative eigenvalues.  There is thus an $SU(2) \times U(1)$ breaking second order phase transition as we vary the parameters.  Near this transition, on the symmetry breaking side, {\it both the size of the symmetry breaking VEV and the mass of the light Higgs field} are much smaller than the nominal multi-TeV scale of the action.  Near the transition, fluctuation corrections also become important and we know\cite{HLM}\cite{CW} that in fact the transition is really a fluctuation induced first order phase transition.  The symmetry breaking VEV jumps to a finite value as the parameters are varied.   To get the actual values of the Z and Higgs boson mass right we have to adjust two parameters, but only one of them has to be finely tuned, to the percent level.   The fields associated with the second Higgs multiplet all get multi TeV scale masses, and the parameter $\tilde{\mu}$ is the mass that pairs up the Weyl fermions in the two Higgs multiplets.   We will discuss gauginos, squarks and sleptons in the next subsection.

\subsection{Standard Model Gaugino Masses}

The confining $SU_P (3)$ gauge theory has composite chiral superfields $M^a_i$ which transform in the adjoint representation of the standard model gauge group.  There are non-zero two point functions
\begin{equation} \langle M^a_i M^b_i \rangle  , \end{equation} and
\begin{equation} \langle D_{\alpha} M^a_i W^a_{\beta i} ,\rangle \end{equation} mixing these fields with the standard model vector field strengths.  The second of these has the scaling properties of a hadron magnetic moment in QCD.  It is proportional to $g_i (\Lambda_3)$ with all other dependence determined by dimensional analysis with no extra powers of $\pi$ .  Since $\Lambda_3$ and the trianon masses are of the same order of magnitude, this mixing will produce Dirac gaugino masses, once we plug in non-zero $F$ terms for the singlets.  \begin{equation} m^{1/2}_i = C_i g_i (\Lambda_3) \sqrt{F} , \end{equation}
where we expect the $C_i$ to be of order $1$. 
The idea is that, treating the $y_{ip}$ couplings as numbers of order $1$, 
there will be terms in the effective action of the form
\begin{equation} \int d^4 \theta g_i  f_i (S/\Lambda_3) D_{\alpha}M^i W_{\alpha}^i , \end{equation} for the $SU(i)$ subgroup of the standard model.  We've seen that the F terms of the singlet fields get expectation values of order $\Lambda_3^2$, so this gives Dirac masses of the required order of magnitude. 
  The Majorana masses of the gauginos are much smaller, of order $\frac{\alpha_i F}{4\pi \Lambda_3 } $, with $F \sim \Lambda_3^2$ for reasons explained above.  They arise from one loop diagrams as explained in\cite{iss}.

One loop sparticle masses coming from the Dirac mass are UV convergent, flavor blind,  and of order 
$\sqrt{\frac{\alpha_i }{4\pi }} m^{1/2}_i$.  This is a well known result from the literature on "supersoft" SUSY breaking\cite{supersoft}. The "UV divergent" contribution from the small Majorana mass is much smaller, because the logarithmic divergence is cut off at the low value $\sqrt{F}$.  SUSY breaking in CSB is soft because it comes from diagrams with gravitinos exchanged with the horizon.  The most significant bounds on the scale $\sqrt{F}$ thus come from the experimental bounds on gaugino, squark and slepton masses, which are on the order of $1 - 3$ TeV.  Therefore $\sqrt{F}$ in the multi TeV range is compatible with experiments, modulo unknown strong $SU_P (3)$ "order one" factors.  

\section{Conclusions}

Very general considerations in quantum gravity lead one to the conclusion that SUSY is necessary for asymptotically flat space-time and that the cosmological constant is a model parameter that must be set rather than calculated\cite{hilbertbundles}.  Observation seems to lead to the conclusion that our own universe will asymptote to de Sitter space, with a cosmological horizon radius of about $10^{61}$ in Planck units.  SUGRA teaches us that vanishing c.c. requires not just SUSY, but a discrete\cite{komseib} R symmetry, if the model is to have a one parameter set of dS perturbations.  Those perturbations break SUSY and the R symmetry in tandem, to keep the c.c. small.  The cosmological horizon naturally breaks the chiral R symmetry, giving the gravitino a mass\footnote{Note that in AdS space the reflection of gravitinos from the boundary preserves both SUSY and the R symmetry.}. 

The strategy of CSB is to find an effective field theory implementation of these quantum gravity rules, which is compatible with existing experiments, and the apparent unification of standard model couplings at about $10^{16}$ GeV.  So far, only a single class of models, the Pyramid Schemes, has been found, which may accomplish this task.  One begins with a spectrum compatible with trinification, with extra vector-like {\it trianons} under each of the $SU_i (3)$ of the trinification group, and also transforming like $3 + \bar{3}$ under a new $SU_P (3)$.   In the zero c.c. limit, this gauge theory flows to a conformal fixed line at low energy, generically at a strongly coupled point.  
At this stage of model building, traditional string theory techniques could be extremely useful.  Are there string constructions that lead to trinification plus this extra structure? Can they have the right number of standard model generations, Yukawa and neutrino textures, {\it etc.} to fit experiment?   Do we have a discrete R symmetry with the properties assumed in this paper? All of those things should be relatively unaffected by CSB.   

The other aspect of these models that could be attacked by traditional string theory methods is the question of the origin of the singlets.  Are they remnants of trinification multiplets?  Do they come from some other aspect of string models?  What determines their number and the values of their couplings $y_{ip}$ to the trianons and $\beta_p$ to $H_u H_d$?  These are all $\Lambda = 0$ questions, within the realm of conventional string/M-theory model building.  

The values of the trianon masses $m_i$, the $\mu$ and $\mu_p^2$ parameters, and the degenerate matrix $\gamma_{pq}$ will have to be extracted from a much more quantitative theory of de Sitter space than anything we have at present.  The degeneracy of $\gamma$ is the most bizarre requirement from an effective field theory point of view, but it follows from the SUSY violating properties of dS space.  It is interesting that once one has imposed that condition, the observed standard model symmetry breaking pattern emerges automatically, once we assume that $SU(2) \times U(1)$ is broken.  The model naturally allows for a weakly first order phase transition and we have to fine tune a single parameter in order to obtain the correct value of the electoweak gauge boson masses and the Higgs boson mass, while keeping the rest of the spectrum consistent with LHC bounds.  It is difficult to assess the complex of issues that are known as the "little hierarchy problem" because our parameters come from the mysterious interactions with the horizon and also involve strongly coupled $SU_P (3)$ physics.  

There seem to be three roads to making future progress on these models.  The first is to delve back into traditional string phenomenology to find a model that fits with the requirements of the vanishing $\Lambda$ limit of the Pyramid schemes:  the Pyramid gauge group broken at the proper unification scale to $SU_P(3) \times G_{std} \times R$ where $R$ is the discrete group that forbids all the couplings we've left out of the model.
That model should also explain the origin of the singlets and calculate the couplings $y_{ip}$, $\beta_p$, and the unification scale values of the gauge and Yukawa couplings of the Pyramid scheme.  This model should also explain the origins of the Yukawa couplings of quarks and leptons and of the dimension $5$ operators that give rise to neutrino masses.  All of this indicates a lot of structure in the range $10^{14} - 10^{16}$ GeV, which must be accounted for  by a theory in asymptotically flat space time and explains a lot of the phenomenology of particle physics.

Next, we must find a quantitative solution of the strongly coupled $SU_P (3)$ SUSY gauge theory, with quark masses of order the confinement scale and Yukawa couplings to  $3$ or more chiral singlets.   This is of course a major undertaking, but it is the kind of task we've been contemplating since the 1970s.  

The hardest of the three paths is to make progress on a more quantitative theory of de Sitter space, in order to calculate the parameters $\mu_p^2, \mu, \gamma_{pq}$.  

It is worth stressing again that the implication of our approach to SUSY breaking is that the structure of the neutrino, quark and charged lepton mass matrices, as well as the origin of CP violation, is assumed to be well described by a supersymmetric theory with vanishing c.c., which is the origin of a set of energy scales between $10^{14} - 10^{16}$ GeV.  The high scale is the scale of coupling unification and the lower scales are associated with neutrino masses.  Powers of ratios between them account for the structure of the CKM and neutrino mass matrices.  In particular, CP violation exists, only because of the presence of the lower scale.  In this case, if the Unruh temperature of gravitinos reflected from the dS horizon is $> 10^{16}$ GeV, we have a natural solution of the strong CP problem. 
This picture implies that some of the tools of old fashioned string phenomenology might be useful in describing the  real world.  They can determine the supersymmetric limit to which our universe would tend if we could tune the c.c. to zero.   

\section{Acknowledgements}

We thank Profs. H. Haber and particularly S. Thomas for useful conversations about SUSY phenomenology and the proper treatment of the low energy Higgs sector.  This paper was completed while the author was attending the workshop Gravity: New Perspectives from Strings and Higher Dimensions, at the Centro de Ciencias Pedro Pascual, Benasque Spain.  The author thanks, R. Emparan,V. Hubeny, M. Rangamani, the organizers, and the staff of the center, for providing a venue where a lot of exciting scientific work could get done. This work was supported in part by the U.S. Dept. of Energy under Grant DE-SC0010008.

\end{document}